\documentclass[floatfix,aps,prl,showpacs,amsmath,nofootinbib,
preprintnumbers,twocolumn]{revtex4}
\usepackage{graphicx}

\def\beqra{\begin{eqnarray}}
\def\eeqra{\end{eqnarray}}
\def\beq{\begin{equation}}
\def\eeq{\end{equation}}

\def\L{\Lambda}
\def\l{\lambda}

\def\bk{{\bf k}}
\def\vp{\varphi}

\def\bp{{\bf{p}}}
\def\bq{{\bf{q}}}

\def\half{\mbox{\small$\frac{1}{2}$}}

\def\agt{\stackrel{>}{\sim}}

\begin{document}

%\preprint{...}

\title{Baryonic Acoustic Oscillations via the Renormalization Group}

\author{Sabino Matarrese}\email{sabino.matarrese@pd.infn.it}
\affiliation{Dipartimento di Fisica ``G.\ Galilei,'' Universit\`{a} di Padova, 
        and INFN, Sezione di Padova, via Marzolo 8, Padova I-35131, Italy}

\author{Massimo Pietroni}\email{massimo.pietroni@pd.infn.it}
\affiliation{INFN, Sezione di Padova, via Marzolo 8, I-35131, Italy}

\date{\today}

\begin{abstract}
A semi-analytic approach to the computation of the non-linear power-spectrum
of dark matter density fluctuations is proposed. 
The method is based on the Renormalization Group technique and can be 
applied to any underlying cosmological model. 
Our prediction on the baryonic acoustic oscillations in a $\Lambda$CDM model   
accurately fits the results of N-body simulations down to zero redshift, 
where perturbation theory fails. 

\end{abstract}


\maketitle

%%%%%%%%%%%%%%%%%%%%%%%%%%%%%%%%%%%%%%%%
%%%%%%%%%%%%%%%%%%%%%%%%%%%%%%%%%%%%%%%%

Cosmological linear perturbation theory is a fundamental tool in 
the physics of the Cosmic Microwave Background (CMB), allowing a careful 
extraction of cosmological parameters from the data.
In the case of Large-Scale Structure, the accuracy of theoretical predictions 
is not yet at the same level than for CMB, due to the higher
degree of non-linearity of the underlying density fluctuations. 
For instance, the study of Baryonic Acoustic Oscillations (BAO) -- 
a fundamental probe for  Dark Energy measurements -- requires a precision 
of a few percent in the theoretical predictions for the matter power-spectrum 
(PS) in the wavenumber range $k\simeq 0.05 - 0.25$ h/Mpc \cite{Komatsu}. 
Including higher orders in PT \cite{review} is known to give a poor 
performance in this range, leaving N-body simulations as the only viable 
approach to the problem.

Recently, however, perturbation theory (PT) has experienced a 
renewed interest, mainly motivated by two reasons. First, next generation
galaxy  surveys are going to measure the PS at 
large redshift, where the fluctuations are still in the linear regime and 
1-loop PT is expected to work \cite{Komatsu}. Second,
Crocce and Scoccimarro \cite{Crocce} have shown that the perturbative 
expansion can be reorganized in a very convenient way, which allows the 
use of standard tools of field theory, like the Feynman diagrams. They 
managed to compute the two-point correlator between density or velocity 
field fluctuations at different times (the `propagator') by  resumming an 
infinite class of diagrams at all orders in PT.  Other approaches 
can be found in Ref.~\cite{RGLSS}.

In this {\it Letter}, we will proceed along the same path, by implementing
Wilsonian Renormalization Group (RG) techniques to compute the PS. 
RG methods, widely used in statistical 
mechanics and quantum field theory 
\cite{RGreview}, are particularly suited to physical situations in 
which there is a separation between the  scale where one is supposed to 
control the `fundamental' theory and the scale were measurements are 
actually made. Starting from the fundamental scale, the RG flow 
 describes the gradual inclusion of fluctuations at scales closer and 
closer to the one relevant to measurements.  The new fluctuations 
which are included at an intermediate step, feel an effective theory,
which has been `dressed' by the fluctuations already included. 
In the present case, the RG flow will start from small wavenumbers 
$k$, where linear theory works, to
reach higher and higher $k$. 

We will consider a self-gravitating system of Dark Matter (DM) particles 
which,  in the ``single-stream'' approximation, is governed by the 
continuity, $\partial_\tau\,\delta + 
{\bf \nabla}\cdot\left[(1+\delta) {\bf v} \right]=0$, and Euler equations 
$\partial_\tau\,{\bf v}+{\cal H} {\bf v} + ( {\bf v} \cdot {\bf \nabla})  
{\bf v}= - {\bf \nabla} \phi$. 
Here $\tau$ is the conformal time, $\delta$ the mass-density fluctuation, 
$\bf v$ the peculiar velocity, 
$\phi$ the peculiar gravitational potential which, on subhorizon scales, 
obeys the Poisson equation $\nabla^2 \phi = \frac{3}{2}
\,{\cal H}^2  \,\delta\,$, with ${\cal H}= d \ln a/d \tau$,
having assumed an Einstein-de Sitter background cosmology.

Following Ref. \cite{Crocce} we 
introduce the doublet $\vp_a({\bf k},\eta)$ ($a=1,2$) 
-- defined in Fourier space -- 
given by
$\left(\varphi_1, \varphi_2\right) \equiv e^{-\eta} \left(\delta, 
- i {\bf k} \cdot {\bf v}/{\cal H}\right)$, 
where the time variable has been replaced by the logarithm of the scale factor,
$\eta= \ln(a/a_{in})$, $a_{in}$ being the scale factor at a conveniently 
remote epoch, when all the relevant scales are well in the linear regime. 
Notice that, compared with the definition in 
Ref.~\cite{Crocce}, we have an overall factor $e^{-\eta}$, 
such that the linear growing mode corresponds to $\vp_a = \mathrm{const}$.
We can then rewrite the continuity and Euler equations in 
compact form, as 
\beqra
\left(\delta_{ab} \partial_\eta+\Omega_{ab}\right) \varphi_b({\bf k}, \eta) 
& = & e^\eta \gamma_{abc}({\bf k},\,-{\bf p},\,-{\bf q}) \nonumber \\
& \times & \varphi_b({\bf p}, \eta )\,\varphi_c({\bf q}, \eta )\,,
\label{compact}
\eeqra
where ${\bf \Omega} = \left(\begin{array}{cc} 1 & -1\\ -3/2 & 3/2
\end{array}\right)\,$ and repeated indices/momenta are summed/integrated 
over.

We have defined a {\it vertex} function, 
$\gamma_{abc}({\bf k},{\bf p},{\bf q})$ ($a,b,c,=1,2$) 
whose only non-vanishing elements are
$\gamma_{121}({\bf k},\,{\bf p},\,{\bf q}) = 
1/2 \,\delta_D ({\bf k}+{\bf p}+{\bf q})\, \alpha(\bp,\bq) $, 
$\gamma_{222}({\bf k},\,{\bf p},\,{\bf q}) = 
\delta_D ({\bf k}+{\bf p}+{\bf q})\, \beta(\bp,\bq)$,
%\beqra
%&&\gamma_{121}({\bf k},\,{\bf p},\,{\bf q}) = 
%\frac{1}{2}\delta_D ({\bf k}+{\bf p}+{\bf q})\, \alpha(\bp,\bq) \,, 
%\nonumber\\
%&&\gamma_{222}({\bf k},\,{\bf p},\,{\bf q}) = 
%\delta_D ({\bf k}+{\bf p}+{\bf q})\, \beta(\bp,\bq)\,, 
%\label{vertice}
%\eeqra
and 
$\gamma_{121}({\bf k},\,{\bf p},\,{\bf q})  = 
\gamma_{112}({\bf k},\,{\bf q},\,{\bf p}) $, where
 $\alpha(\bp,\bq )= [(\bp + \bq) \cdot \bp]/p^2$ and
$\beta(\bp,\bq ) = (\bp + \bq)^2 
(\bp \cdot \bq)/(2 p ^2 q^2)$. 

Besides the vertex, the other relevant dynamical quantity is the linear 
retarded {\it propagator}, which gives the linear 
evolution in $\eta$ of $\vp_a$, $\varphi_a^0( {\bf k},\eta) = 
g_{ab}(\eta,\eta^\prime)  \varphi_b^0({\bf k},\eta^\prime)$, with 
$\eta>\eta^\prime$ (the subscript $``0"$ indicates solutions of 
the linear equations, $e^\eta \gamma_{abc} =0$), which reads 
$g_{ab}(\eta,\eta^\prime) = [ {\bf B} + {\bf A}\, \exp(-5/2 
(\eta -\eta^\prime))]_{ab}\, \theta(\eta-\eta^\prime)$, with $\theta$ the 
step-function, 
${\bf B} = \frac{1}{5}\left(\begin{array}{cc}
3 & 2\\ 3 & 2 \end{array}\right)$ and 
${\bf A} = \frac{1}{5}\left(\begin{array}{rr}
2 & -2\\-3 & 3 \end{array}\right)$. 
The growing ($\vp_a \propto \mathrm{const.}$) and the decaying 
($\vp _a\propto \exp(-5/2 \eta)$) modes can be selected by 
considering initial fields $\vp_a$ proportional to 
$u_a = (1,1)$ and $v_a=(1, -3/2)$, respectively. 

To extend the validity of this approach to $\Lambda$CDM, we 
reinterpret the variable $\eta$ as the logarithm of the linear
growth factor of the growing mode, i.e. 
$\eta=\ln (D^+/D^+_{in})$. This approximation has been shown to 
accurately fit N-body simulations for different cosmologies \cite{Komatsu}.

Our aim is to apply methods familiar in quantum field theory to construct 
generating functionals for quantities like the PS, bispectrum, etc. ...
The starting point is to write down an action giving the equation of motion 
(\ref{compact}) at its extrema. One can realize that a new, auxiliary, 
doublet field $\chi_a$ has to be introduced to this aim, and that the action 
is given by 
\beqra
 \label{action}
&&
S= \int d\eta \left[ \chi_a(-\bk,\eta)\left(\delta_{ab} 
\partial_\eta+\Omega_{ab}\right) \varphi_b( {\bf k},\eta) \right.
\\
&& \left. - e^\eta \gamma_{abc}(-\bk, -\bp,-\bq)
\chi_a(\bk,\eta)\varphi_b(\bp,\eta)
\varphi_c( \bq ,\eta) \right] \,. \nonumber
\eeqra

Varying the action $S$ w.r.t. $\chi_a$ gives precisely Eq.~(\ref{compact}), 
while varying it w.r.t. $\vp_a$ yields an equation solved by $\chi_a=0$. 

The probability of a classical field configuration is a delta function 
centered on the solution of the equation of motion. 
By averaging over Gaussian initial conditions, one can show \cite{ms2} 
that the generating functional of correlation functions involving the 
$\vp$ and $\chi$ fields is the path-integral 
\beqra
\label{genf}
&& Z[J_a,\, K_b;\,P^0]  \\
&=&\int {\cal D}  \vp_a {\cal D} \chi_b
\exp \biggl\{\int d\eta d\eta' \left[ -\half \chi_a P^0_{ab} 
\delta(\eta) \delta(\eta^\prime)\chi_b \right. \nonumber\\
&+& \left.
i \chi_a g^{-1}_{ab} \vp_b  - i e^\eta\,\gamma_{abc} \chi_a \vp_b \vp_c + 
i J_a \vp_a + i K_b \chi_b\right]\biggr\} \,, \nonumber
\eeqra
where $J_a$ and $K_b$ are external sources and the 
momentum dependence is everywhere implicit. Here $P^0_{ab}(k) = 
P^0(k) u_a u_b$ and $P^0$ is the linear PS at the 
initial ``time'' $\eta=0$, having assumed pure growing-mode initial conditions.
As usual, we define also the generator of connected Green's functions 
$W[J_a,K_b]=-i \ln Z[J_a,K_b]$ and, for 1-particle irreducible (1PI)
ones, 
the effective action $\Gamma[\vp_a,\, \chi_b] =  W[J_a, K_b] - 
\int d\eta\,d^3 \bk (J_a \vp_a + K_b \chi_b )$, in terms of the ``classical'' 
fields, $\vp_a[J_c, K_d] = \delta W[J_c, K_d]/ \delta J_a$ and 
$\chi_b[J_c, K_d] = \delta W[J_c, K_d]/ \delta K_b$. 
In linear theory the path-integral can be 
performed analytically,  
\beqra
Z_0[J_a,K_b;P^0] &=& \exp \biggl\{- \int d\eta d\eta^\prime
 [\half J_a(\eta) 
P^L_{ab}(\eta,\eta^\prime) J_b(\eta^\prime)
\nonumber \\
& +&i J_a(\eta) g_{ab}(\eta,\eta^\prime) 
K_b(\eta^\prime)] \biggr\}\,,
\label{zfree}
\eeqra
with  $P^L_{ab}(\eta_a,\eta_b; k) = g_{ac}(\eta_a,0)
g_{bd}(\eta_b,0) P^0_{cd}(k)$ the linearly evolved PS.

\begin{figure}
\centerline{\includegraphics[width = 2.2in , bb= 156 80 840 710, 
keepaspectratio=true
]{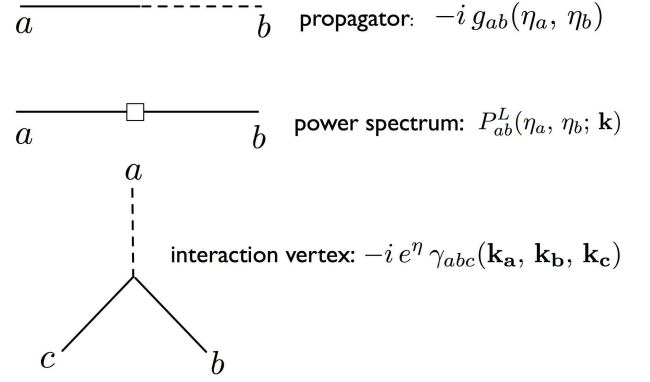}}
\caption{Feynman rules}
\label{Feynman}
\end{figure}

Standard methods of PT can be applied, 
using the Feynman rules sketched in Fig.~\ref{Feynman}. 
There are three building blocks: the linear propagator $g_{ab}$, the linear
PS $P^L_{ab}$ and the vertex $e^\eta\gamma_{abc}$. 
The Feynman diagrams constructed using these rules are in 
one-to-one correspondence with those of Ref.~\cite{Crocce}, and 
reproduce all the known results of PT \cite{review}.

The full propagator, PS and vertex can be obtained from $W$ 
and $\Gamma$ as $(\delta^2 W/\delta J_a \delta K_b)_{J,\,K=0} \equiv - 
\delta({\bk}+{\bk}') G_{ab}$, $(\delta^2 W/\delta J_a 
\delta J_b)_{J,\,K=0} \equiv i \delta({\bk}+{\bk}') P_{ab}$ and 
$(\delta^3 \Gamma/\delta \chi_a \delta \vp_b \delta \vp_c)_{\vp,\,\chi=0} 
\equiv - \Gamma_{abc}(\bk_a+\bk_b + \bk_c;\eta)$.  

Using the definitions of effective action and classical fields above, 
one can show that the full PS has the structure
$P_{ab}=P^I_{ab}+P^{II}_{ab}$, where 
$P^I_{ab}(k; \eta, \eta^\prime) = G_{ac}(k;\eta,0)
G_{bd}(k;\eta^\prime,0) P^0_{cd}(k)$ and 
$P^{II}_{ab}(k; \eta, \eta^\prime)= \int_0^\eta d\eta^{\prime\prime}
\int_0^{\eta^\prime} d\eta^{\prime\prime\prime}
G_{ac}(k;\eta,\eta^{\prime\prime})
G_{bd}(k;\eta^\prime,\eta^{\prime\prime\prime}) 
\Phi_{cd}(k;\eta^{\prime\prime},\eta^{\prime\prime\prime})$, where $\Phi_{ab}$ 
is defined through 
$(\delta^2 \Gamma/\delta \chi_a \chi_b)_{\vp,\chi=0} 
\equiv i  P^0_{ab}(k) \delta(\eta) \delta(\eta^\prime) + i \Phi_{ab}$. \\

The starting point of our formulation of the RG is a modification of 
the primordial PS appearing in Eq.~(\ref{genf}), 
as $P^0(k) \rightarrow P^0_\l (k) = P^0(k) \theta(\l-\,k)$.
Inserting the truncated PS in Eq.~(\ref{genf}) yields a 
generating functional $Z_\l[J_a,\, K_b;\,P^0] 
\equiv Z[J_a,\, K_b;\,P^0_\l]$, that describes a fictitious Universe, 
whose statistics of initial data  
is modified by suppressing all fluctuations with wavenumber larger than $\l$. 
On the other hand, the dynamical content, encoded in the linear propagator 
and in the structure of the interaction, is left unchanged. 
 
In the $\l \to \infty$ limit, all the fluctuations are included, and we 
recover the physical situation. Increasing the cutoff from $\l=0$ to 
$\l\to \infty$, the linear and non-linear effect of fluctuations of higher 
and higher wavenumber is gradually taken into account. 
This process is described by a RG equation \cite{RGreview, ms2}
 which can be obtained by 
taking the $\l$ derivative of $Z_\l$,
\beq
\partial_\l Z_\l = 
\frac{1}{2} \int d\eta_a\,d\eta_b d^3\bq\,\delta(\l-q) 
P^0_{ab}(q) \delta(\eta_a) \delta(\eta_b)
\frac{\delta^2 Z_\l }{\delta K_b \delta K_a} \,.
\label{ZRG}
\eeq

\begin{figure}
\centerline{\includegraphics[width = 2.7in ,  bb= 70 30 540 150,
%keepaspectratio=true
]{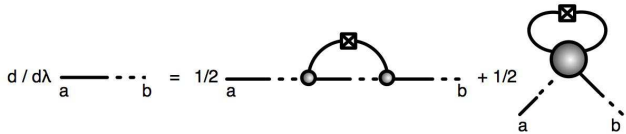}}
\caption{RG equation for the propagator $G_{ab,\l}$}
\label{prop_RG}
\end{figure}

\begin{figure}
\centerline{\includegraphics[width = 2.7 in ,  bb= 70 30 540 200
keepaspectratio=true
]{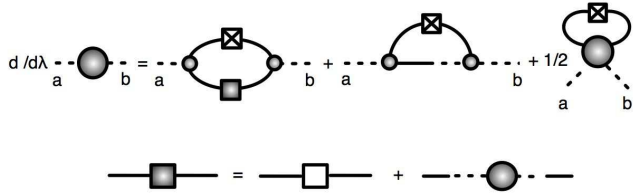}}
\caption{RG equation for $\Phi_{ab,\l}$}
\label{runPhi}
\end{figure}

Taking successive derivatives w.r.t. the sources of this master equation
 -- or of the  analogous ones for $W$ and $\Gamma$ --   
one can obtain the RG evolution of any physical quantity. 
The structure of the equation is such that the evolution of 
a correlation function of order $n$ involves all correlations up 
to order $n+2$. 
The RG equation for the propagator and for the function $\Phi_{ab,\l}$
are represented in Figs.~\ref{prop_RG} and \ref{runPhi}, respectively, 
where the thick lines indicate full
propagators, the dark box is the full PS $P_{ab,\l}$, 
dark blobs are full 1PI functions, and the 
crossed box is the RG kernel obtained by deriving w.r.t. $\l$ the $\theta$ 
function multiplying the PS in $P^I$, that is
\beq
\label{kernel}
K_{ab,\l}(k,\eta,\eta^\prime) = G_{ac, \l}(k;\eta,0)
G_{bd, \l}(k;\eta^\prime,0) P^0_{cd}(k)\,\delta(\l-k) \;. 
\eeq


A recipe can be given to obtain the RG equation for any given 
quantity \cite{ms2}: 

{\it i)} write down the {\em 1-loop expression} 
for the quantity of interest, obtained using  {\em any needed vertex},
(for instance, in Figs. \ref{prop_RG},
\ref{runPhi} we have not only the vertex $\chi \vp \vp$, but 
also $\chi\chi\vp$,
$\chi \vp\vp\vp$ and $\chi \chi\vp\vp$, which vanish at tree-level);  

{\it ii)} promote the linear propagator, the PS and the 
vertices appearing in 
that expression to full, $\l$-dependent ones;

{\it iii)} take the $\l$-derivative of the full expression, 
by considering only 
the explicit $\l$-dependence of the step-function contained in 
$P^I_\l$.

It should be emphasized that the RG equations obtained following 
these rules are {\em exact}, in the sense that they encode all the 
dynamical and statistical content of the path-integral (\ref{genf}) 
or, equivalently, 
of the continuity and Euler equations supplemented by the initial PS. 
Different approximation schemes can be attempted, including of 
course PT, which can be recovered by using quantities up to
the $l$-th loop order in the RHS to get the $l+1$-th one by 
performing the $\l$-integration.
However, our RG equations are most indicated for non-perturbative 
resummations. 

As a first step, we note that the full 3 and 4-point functions 
appearing in the RG equations for $G_{ab,\l}$ and $\Phi_{ab,\l}$ are
also $\l$-dependent quantities, which can be computed by RG equations, 
also derived from Eq.~(\ref{ZRG}). These equations depends, in turn, 
on full, connected, and $\l$-dependent functions up to 5 
(for the 3-point function) or 6 
(for 4-point ones) external legs, which also evolve according 
to RG equations.  Approximations to the full RG flow then amount
to truncating the full hierarchy of coupled differential equations, and 
using some ansatz for the full n-point functions appearing in the
surviving equations. We will approximate the full RG flow by keeping
the running of the 2-point functions (propagator and PS) and
keeping the tree-level expression for the trilinear vertex $\chi\vp\vp$. 
In this approximation, only the first diagrams on the RHS of Figs. 2 and 3 
contribute to the running.    

The 1-loop result for the propagator \cite{review} can be recovered 
by using the tree-level expressions for the kernel $K_{ab,\l}$ 
and for the propagators on the RHS of Fig. 2. 

Using running propagators on the RHS of Fig. 2, while keeping the kernel at  
tree-level, we get a RG equation which can be analytically integrated in the 
$k\gg q=\l$ limit, 
\beqra
\partial_\l G_{ab,\l} (k;\eta_a,\eta_b)&=&- G_{ab,\l} (k;\eta_a,\eta_b) \, 
\frac{k^2}{3} \,
 \frac{\left(e^{\eta_a} -e^{\eta_b}\right)^2}{2} 
 \,\nonumber\\
&& \times \int d^3\bq
\,\delta(\l-q) \,\frac{P^0(q)}{q^2}\,,
\eeqra
having used the property
$u_f \gamma_{efg}(-\bk,\bq,\bk-\bq) \simeq \delta_{eg} 
(k/2q) \cos \bk\cdot\bq$, valid in this limit. 
Imposing the initial condition $G_{ab,\l=0} ( k,\eta_a,\eta_b) = 
g_{ab} (\eta_a,\eta_b)$, and integrating up to $\l=\infty$, one gets  
\beq G_{ab} (k,\,\eta_a,\,\eta_b) =g_{ab}(\eta_a,\,\eta_b) 
e^{ - k^2 \sigma_v^2 \,
 \frac{\left(e^{\eta_a} -e^{\eta_b}\right)^2}{2} }\,,
 \label{resu}
\eeq
where $\sigma_v^2$ is the velocity dispersion, defined as
$\sigma_v^2 \equiv (1/3) \int d^3\bq P^0(q)/q^2$. Two comments are in 
order. The RG improvement 
discussed here has a clear interpretation in terms of PT. 
Indeed, as shown in \cite{Crocce}, the result in Eq.~(\ref{resu}) 
can be obtained also by resumming an infinite class of diagrams.
It is amazing how the same result, that in PT requires
a careful control of the combinatorics, is here obtained by a simple, 
1-loop, integration. 
The second comment has to do with the dramatic modification of the UV
behavior of the resummed propagator w.r.t. 
the linear one, and on its impact on the RG flow. 
Indeed, when the propagator of Eq.~(\ref{resu}) is employed in the kernel 
$K_{ab,\l}$, an intrinsic UV cut-off is provided to the RG flow: fluctuations 
with large momenta are exponentially damped, so that the RG 
evolution freezes out for $\l \gg e^{-\eta} /\sigma_v$. 
This is a genuinely non-perturbative effect, 
which is masked if one considers PT at any finite order. 
In other words, the full equations of motion are much better behaved 
in the UV than their perturbative approximations. 

Improving further on the approximations leading to Eq.~(\ref{resu}), 
we relax the $k\gg \l$ condition, keeping the full momentum
dependence of the tree-level vertex, and using $\l$-dependent
propagators also in the kernel (\ref{kernel}).
For the full propagator we use the ansatz $
G_{ae,\l}(k,\,s_1,\,s_2) = H_{ae,\l}(k) \,\exp[ - k^2 \sigma_v^2 \,
 \frac{\left(e^{s_1} -e^{s_2}\right)^2}{2} ]$,
in which we have factored out the leading time behavior of Eq.~(\ref{resu}). 
Inserting the expression above in the RG equations we obtain a closed 
system of four coupled differential equations, one for 
each component of $H_{ae,\,\l}$. In order to simplify it further, we 
consider the two combinations $G_{a1,\l}+G_{a2,\l}$,
($a=1,\,2$), therefore, for each external momentum $k$, we solve
a system of two coupled 
differential equations. Full details will be given in \cite{ms2}. 
The exponential damping of Eq.~(\ref{resu}) is exhibited also by our 
more refined scheme. 

The RG evolution of the propagator governs that of the $P^I_{ab,\l}$ 
contribution to the full PS. On the other hand, the evolution of the 
other contribution, $P^{II}_{ab,\l}$, can be computed by solving the 
equation in Fig.~\ref{runPhi}. We approximate the exact RG equations 
along the same lines we follow in the computation of the propagator. 
In particular, the $\eta$ integrations in the expression for 
$P^{II}_{ab}$ require an ansatz for its `time' dependence. In analogy 
with Eq.~(\ref{resu}), we use, again on the RHS of the RG equation,
$P^{II}_{ab,\l}(q, \,\eta_a, \,\eta_b) = \bar{P}^{II}_{ab,\l}(q, 
\,s_1, \,s_2)\,\exp[ - q^2 \sigma_v^2 \,
\frac{\left(e^{\eta_a} -e^{s_1}\right)^2 + 
\left(e^{\eta_b}-e^{s_2}\right)^2}{2}]$. We compute the PS at equal `times',  
$\eta_a=\eta_b=\eta$, with the initial condition 
$P^{II}_{ab,\l=0}(k) = 0\,$. We consider a spatially flat $\L$CDM model 
with $\Omega_\L^0=0.7$, $\Omega_b^0=0.046$, $h=0.72$, $n_s=1$. 
The primordial PS $P^0$ is taken from the output of linear theory at 
$z_{in}=35$, as given by the CAMB Boltzmann code \cite{CAMB}.
\begin{figure}
\centerline{\includegraphics[width = 3.5in,  bb= 0 20 1050 780,
%keepaspectratio=true
]{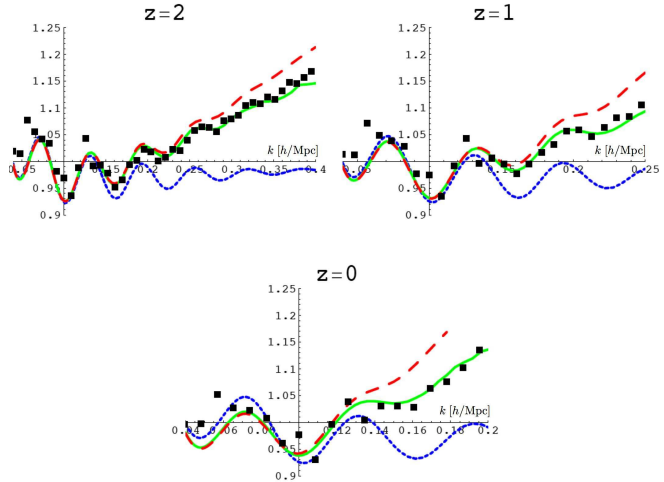}}
\caption{The power-spectrum at $z=2,\,1,\,0$, as given by the RG 
(solid line), linear theory (short-dashed), 1-loop PT (long-dashed), 
and the N-body simulations of \cite{White} (squares).}
\label{Pplot}
\end{figure}

In Fig.~\ref{Pplot} we plot our results for the PS (solid lines), in the
momentum range relevant for the BAO, at $z=0,\,1$, and $2$. 
The short-dashed lines correspond to the linear theory and the 
long-dashed ones to 1-loop PT (which, at $z=0$ has been truncated for $k\agt 
0.17$ h/Mpc, where $P^{I}$ takes 
negative values, signaling the breakdown of the perturbative expansion). 
The black squares are taken from the numerical
simulations of Ref.~\cite{White}. To enhance the BAO feature, each 
PS has been divided by the linear one, in a model without baryons
\cite{Eisenstein}.
In the peak region, our RG results agree with those of
N-body simulations to a few percent accuracy
down to redshift $z=0$, where linear and 1-loop
perturbation theory badly fail. Thus, the dynamical behavior in this
momentum range appears to be captured fairly well by the approximations 
implemented by our approach, namely the `single stream approximation', 
leading to Eq.~(\ref{compact}), and the non-linear corrections
of the two-point functions only, i.e. the propagator and the PS. 
The RG performance can be systematically improved by 
increasing the level of truncation of the full tower of differential 
equations, the next step being the inclusion of the running of the 
trilinear vertex.

We thank M. White, for providing us with the N-body data of 
Ref.~\cite{White}, N. Bartolo, P. McDonald and M. Viel, for discussions. 
M.P. thanks the Galileo Galilei Institute for Theoretical Physics 
for hospitality during the initial stages of this work.

%%%%%%%%%%%%%%%%%%%%%%%%%%%%%%%%%%%%%%%%
%%%%%%%%%%%%%%%%%%%%%%%%%%%%%%%%%%%%%%%%

%%%%%%%%%%%%%%%%%%%%%%%%%%%%%%%%%%%%%%%%
%%%%%%%%%%%%%%%%%%%%%%%%%%%%%%%%%%%%%%%%

%%%%%%%%%%%%%%%%%%%%%%%%%%%%%%%%%%%%%%%%
%%%%%%%%%%%%%%%%%%%%%%%%%%%%%%%%%%%%%%%%
\end{document}